\documentclass[prl,twocolumn,groupedaddress]{revtex4}
\newcommand{\Rv}{\vec{R}}

\newcommand{\xv}{{\bf x}}

\newcommand{\uv}{\vec{u}}

\newcommand{\qv}{{\bf q}}

\def\mm#1{\underline{\underline{{#1}}}}

\newcommand{\Tr}{{\rm Tr}}

\newcommand{\ppi}{{\partial_i}}
\newcommand{\ppj}{{\partial_j}}
\newcommand{\pz}{{\partial_z}}

\usepackage[dvips]{graphicx}

\begin{document}
\title[]{Thermal fluctuations and anomalous elasticity of homogeneous
  nematic elastomers} 
\author{Xiangjun Xing} 
\author{Leo Radzihovsky}
\affiliation{Department of Physics, University of Colorado,
   Boulder, CO 80309}

\date{\today}

\begin{abstract}
  We present a unified formulation of a rotationally invariant {\em
    nonlinear} elasticity for a variety of spontaneously anisotropic
  phases, and use it to study thermal fluctuations in nematic
  elastomers and spontaneously anisotropic gels. We find that in a
  thermodynamic limit homogeneous nematic elastomers are universally
  incompressible, are characterized by a universal ratio of shear
  moduli, and exhibit an anomalous elasticity controlled by a
  nontrivial low temperature fixed point perturbative in
  $D=3-\epsilon$ dimensions.  In three dimensions, we make predictions
  that are asymptotically exact.

\end{abstract}

\maketitle

\begin{figure}[!htbp]
\begin{center}
  \includegraphics[]{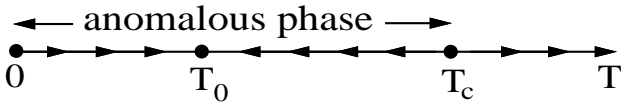}
\vspace{-4mm}
\caption{Renormalization group flow for anomalously elastic
  solids, with $T_c$ the transition temperature to the ordered state
  and $T_o$ a nontrivial infrared stable fixed point controlling
  properties of the strongly interacting ordered, critical state.}
\label{flow_figure}
\end{center}
\vspace{-6mm}
\end{figure}

Qualitatively important effects of fluctuations are generically
confined to the vicinity of isolated critical points, where a system is
tuned to be ``soft'', characterized by low energy excitations. As long
as the ordered state is stable, fluctuations about it are typically
described by a harmonic theory, controlled by a Gaussian fixed point.
However, there exists a novel class of systems, that includes
smectic\cite{GP,RT_smectic} and columnar liquid
crystals\cite{RT_columnar}, vortex lattices in putative magnetic
superconductors\cite{RT_MSC}, polymerized membranes\cite{membranes},
and nematic elastomers\cite{GL,review_elastomers,LMRX,XRdisorderPRL},
whose ordered states are a striking exception to this rule. A unifying
feature of these phases is their underlying, spontaneously broken
rotational invariance, that strictly enforces a particular
``softness'' of the corresponding Goldstone mode Hamiltonian. As a
consequence, the usually small nonlinear elastic terms are in fact
comparable to harmonic ones, and therefore must be taken into account.
Similarly to their effects near continuous phase transitions, but
extending throughout an ordered phase, fluctuations drive
nonlinearities to qualitatively modify such soft states.  The
resulting strongly interacting ordered states at long scales exhibit
rich phenomenology such as a universal nonlocal elasticity, a strictly
nonlinear response to an arbitrarily weak perturbation and a universal
ratio of wavevector-dependent singular elastic moduli, all controlled
by nontrivial infrared stable fixed points illustrated in
Fig.\ref{flow_figure}.

Nematic elastomers and gels are weakly crosslinked polymer-liquid
crystal composite materials, which, when nematically ordered exhibit
fascinating elastic and electro-optic properties, and are therefore of
considerable technological importance.  The most unusual of these is
the divergent strain response to a stress applied transversely to the
spontaneous nematic direction\cite{softmodes,Olmsted}, an elastic
softness that is a consequence of the novel nemato-elastic Goldstone
mode corresponding to a spontaneous orientational ordering and an
accompanying elastic distortion of the
network\cite{GL,LMRX,semisoft}. Despite of considerable progress in
understanding the properties of these liquid crystalline
rubbers\cite{review_elastomers}, it was only recently appreciated,
that (because of their harmonic softness) nematic elastomers and gels
belong to the aforementioned class of anomalously elastic materials,
and are the first example of three-dimensional solids with such novel
fluctuation-driven properties.

In this Letter we present a unified formalism for a systematic
derivation of the rotationally-invariant nonlinear elastic energy for
variety of soft elastic media. This is a nontrivial ingredient that is
a prerequisite to treating effects of fluctuations and external
perturbations in such soft materials.  We utilize this formalism to
study long-scale elastic properties of homogeneous fluctuating nematic
elastomers, and demonstrate that they indeed exhibit anomalous
elasticity phenomenology discussed above\cite{Stenull-Lubensky}.

A configuration of an elastic media is described by a mapping $\xv
\mapsto \Rv(\xv)$ from the $D$-dimensional reference space $\xv$,
which labels mass points of the media, to the $d$-dimensional target
space $\Rv$. Our goal is to construct a {\em nonlinear} elastic energy
functional $\cal H$, describing long-scale phonon deformations
$\uv(\xv)=\Rv(\xv)-\Rv^0(\xv)$ about the spontaneously ordered state
$\Rv^0(\xv)$, with $\cal H$ exhibiting all rotational and
translational symmetries of the reference and target spaces
\cite{LMRX}. This is nontrivial to implement at a nonlinear level as
some of the symmetries are spontaneously broken by the ground state,
and are therefore not manifest (hidden) in $\cal H$.

A systematic way of constructing such $\cal H$ is by expanding it in
in terms of {\em scalar} nonlinear operators $S_n$
\begin{eqnarray}
{\cal H}[\mm{g}] &=& {\cal H}[\mm{g}_0] + a_{n} S_n
 + \frac{1}{2} a_{n m} S_{n} S_{m} + \ldots\,\label{expansion}\\
S_n &=& \Tr (\mm{g}^n - \mm{g}_0^n)\;.\hspace{3mm}
n = 1, 2, \ldots, {\mbox {Min(d,D)}},\label{S_invariants}
\end{eqnarray}
with $g_{ij}=\partial_i\Rv\cdot\partial_j\Rv$ and
$g_{ij}^0=\partial_i\Rv^0\cdot\partial_j\Rv^0$ metric tensors
corresponding to $\Rv(\xv)$ and $\Rv^0(\xv)$ configurations and
summation convention over repeated indices is used throughout.
Eliminating the metric tensor
\begin{equation}
g_{ij}=g_{ij}^0 + 2{\partial R^0_\alpha\over\partial x_i}
u_{\alpha\beta}{\partial R^0_\beta\over\partial x_j}\label{delta_g}
\end{equation}
in favor of the Lagrangian strain tensor
\begin{eqnarray}
u_{\alpha\beta}&=&{1\over2}\left({\partial\Rv\over\partial
R^0_\alpha}\cdot{\partial\Rv\over\partial
R^0_\beta}-\delta_{\alpha\beta}\right),\\
&=&{1\over2}(\partial_\alpha u_\beta+\partial_\beta u_\alpha +
\partial_\alpha\uv\cdot\partial_\beta\uv)
\end{eqnarray}
measured in terms of the ground state coordinate system $R^0_\alpha$
(rather than $x_i$) leads to the desired nonlinear elastic energy
${\cal H}[u_{\alpha\beta}]$.  The necessity of this procedure is that
it ensures that the expansion in $u_{\alpha\beta}$
about the spontaneously symmetry-broken ground state is
rotationally invariant in both the target and reference spaces.

We have used this general formalism, with differences encoded in $D$,
$d$ and the nature of the broken ground state, $\Rv^0(\xv)$, to derive
nonlinear elastic Hamiltonians for a variety of elastic
media\cite{XRunpublished}, some of which have previously appeared in
the literature. In this Letter we apply it to study fluctuations in
the most nontrivial and heretofore unexplored system, the
spontaneously-uniaxial nematic elastomers with
$d=D$\cite{review_elastomers,LMRX}.

As discussed in detail in Refs.\ \onlinecite{GL,LMRX}, most of the
properties of nematic elastomers can be captured by a purely elastic
description in terms an elastic strain tensor, with a nematic order
parameter $Q_{\alpha\beta}=S(\hat{n}_\alpha\hat{n}_\beta -
\delta_{\alpha\beta}/3)$ ($S$ and $\hat{n}$, respectively, the
magnitude and the unit director for nematic order)
integrated out.  The isotropic-nematic transition is then
characterized by a spontaneous uniaxial distortion
$u^0_{\alpha\beta}\propto Q_{\alpha\beta}$, corresponding to a
ground-state conformation and a metric tensor given by
\begin{eqnarray}
\Rv^0(\xv)&=&\zeta_\perp\xv_\perp + \zeta_z z \hat{z},\\
g_{\alpha\beta}^0&=&\zeta_\alpha^2\delta_{\alpha\beta},
\end{eqnarray}
with $\xv_\perp$ a $D-1$-dimensional vector transverse to the nematic
director $\hat{n}=\hat{z}$.

In the physical case of $D=d=3$ there are three nonlinear rotationally
invariant operators $S_n$, which, using Eqs.\ref{S_invariants} and
\ref{delta_g}, can be readily shown to be given by
\begin{eqnarray}
S_1&=&2\zeta_\alpha^2 u_{\alpha\alpha},\label{S1}\\
S_2&=&4\zeta_\alpha^4 u_{\alpha\alpha} + 4\zeta_\alpha^2\zeta_\beta^2
u_{\alpha\beta}u_{\beta\alpha},\label{S2}\\
S_3&=&6\zeta_\alpha^6 u_{\alpha\alpha} + 12\zeta_\alpha^4\zeta_\beta^2
u_{\alpha\beta}u_{\beta\alpha}+8\zeta_\alpha^2\zeta_\beta^2\zeta_\gamma^2
u_{\alpha\beta}u_{\beta\gamma}u_{\gamma\alpha},\hspace{0.7cm}\hfill\label{S3}
\end{eqnarray}

Using these expressions inside the expansion for $\cal H$,
Eq.\ref{expansion}, taken to quadratic order in $S_n$'s, we obtain
\begin{eqnarray}
\hspace{-0.3cm}{\cal H} &=& a_z u_{zz} + a_{\perp} u_{ii}\\
&+&{1\over2}\left(\mu_{zi} u_{zi}^2 + B_z u_{zz}^2 
+ \lambda_{zi}u_{zz}u_{ii} + \lambda u_{ii}u_{jj}
+ 2\mu u_{ij} u_{ij}\right)\nonumber\\
&+&b_1 u_{zz} u_{iz}^2 + b_2 u_{kk}u_{iz}^2 + b_3 u_{ij}u_{iz} u_{jz}
+ c u_{iz}^2 u_{iz}^2\nonumber,\label{Hne_full}
\end{eqnarray}
where Roman indices ($i,j,k,\ldots$) take values in the $\perp$-space
only and 
\begin{eqnarray}
a_{z,\perp} &=& 2a_1\zeta_{z,\perp}^2 + 4a_2 \zeta_{z,\perp}^4 + 
6a_3 \zeta_{z,\perp}^6,\\
\mu_{zi} &=& 16a_2\zeta_z^2\zeta_\perp^2 + 24a_3 (\zeta_\perp^2 +
\zeta_z^2)\zeta_z^2\zeta_\perp^2,\nonumber\\ 
&=&{4\over\zeta_z^2-\zeta_\perp^2}
(\zeta_\perp^2a_z-\zeta_z^2a_{\perp}),\label{coefficients}
\end{eqnarray}
with the detailed form of other elastic constants not important
here\cite{XRunpublished}.

\begin{figure}[!htbp]
\begin{center}
  \includegraphics[width=8cm,height=3.5cm]{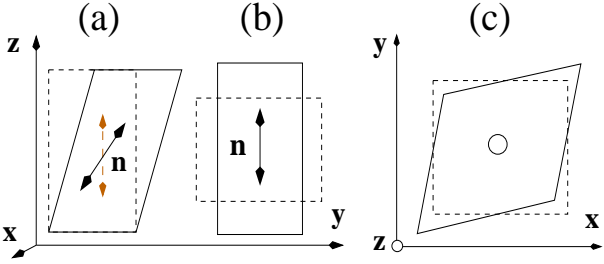}
\caption{(a) simple, (b) longitudinal and (c) transverse shear; in
(c), nematic order is out of the paper.  Simple shear transforms the
system into an equivalent nematic ground state, and thus costs no
energy.}
\label{shear_config}
\end{center}
\vspace{-5mm}
\end{figure}

In the absence of fluctuations, our expansion about the ground state
$\mm{g}^0$, characterized by $\zeta_{z,\perp}$ ensures that $a_z =
a_{\perp}=0$ in equilibrium. Given that in the nematic ground state,
$\zeta_z\neq\zeta_\perp$, this then leads to a strict vanishing of the
shear modulus $\mu_{zi}=0$, as anticipated by Golubovic and
Lubensky\cite{GL,LMRX} and is implicit in the neo-classical theory of
nematic elastomers\cite{review_elastomers,Olmsted}. It corresponds to
a vanishing energy cost of a shear distortion $u_{zi}$ displayed in
Fig.\,\ref{shear_config}$a$, that is one of many fascinating
properties of nematic elastomers and is a manifestation of the
``soft'' Goldstone mode elasticity, discussed in the introduction. At
finite temperature, fluctuations renormalize all elastic constants and
in particular generate nonzero $a_z$, $a_\perp$. However, Ward
identities associated with the underlying rotational invariance of
$\cal H$ ensure that such linear in strain terms can always be
eliminated by shifting to a true, thermally renormalized spontaneous
deformation $\mm{g}^0_R$\cite{XRunpublished}.

Keeping only the most relevant terms in $\cal H$, Eq.\ref{Hne_full},
and using nontrivial relations between elastic
coefficients\cite{XRunpublished}, enforced by the rotational
invariance, we arrive at the effective nonlinear elastic Hamiltonian
for a uniaxial nematic elastomer:
\begin{eqnarray}
{\mathcal H}_{NE} &=& \frac{B}{2} (w_{zz} + w_{ii})^2
 + C (w_{zz}+w_{ii})(w_{zz}-w_{ii})\nonumber\\
&+& \frac{\mu_L}{2} (w_{zz}-w_{ii})^2
+ \mu \tilde{w}_{ij} \tilde{w}_{ij} + 
\frac{K}{2} (\nabla_{\perp}^2 u_z)^2,\hspace{1cm}
\end{eqnarray}
where
\begin{eqnarray}
w_{zz} &=&  \pz u_z + \frac{1}{2} (\nabla_{\perp} u_z)^2,\nonumber\\
w_{ij} &=&   \frac{1}{2} (\ppi u_j + \ppj u_i )
- \frac{1}{2}\ppi u_z \ppj u_z,\nonumber\\
\tilde{w}_{ij} &=& w_{ij} - \frac{w_{kk}}{D-1} \delta_{ij}
\end{eqnarray}
are rotationally-invariant smectic-like and columnar-like nonlinear
strain tensors relative to the uniaxial state,
\begin{eqnarray}
B &=& \frac{1}{4} (B_z + B_{\perp} + \lambda_{zi}),\\ 
\mu_L &=& \frac{1}{4} (B_z + B_{\perp} - \lambda_{zi}),\\ 
C &=& \frac{1}{4} (B_z - B_{\perp}),
\end{eqnarray} 
and $B_\perp=\lambda + 2\mu/(D-1)$ is the $\perp$-plane bulk
modulus. It is easy to see that $B$ is the overall bulk modulus,
$\mu_L$ and $\mu$ are the longitudinal and transverse shear moduli,
respectively corresponding to deformations displayed in
Fig.\,\ref{shear_config} b,c, and $C$ couples bulk mode with
longitudinal shear.  Symmetry requires that transverse shear mode
remains decoupled from other modes.  Because of the vanishing
$\mu_{zi}$, we have also added a curvature $K$ elastic term for $u_z$
to ensure stability at long scales.

Using ${\cal H}_{NE}$, we now proceed to study effects of thermal
fluctuations on nearly homogeneous nematic elastomers.  Standard
analysis leads to the Fourier transform of harmonic correlators
$G^0_{\alpha\beta}(\xv)=\langle u_\alpha(\xv)u_\beta(0)\rangle_0$:
\begin{eqnarray}
G^0_{zz}(\qv) &=& \frac{1}{B_z (1 - \rho)q_z^2 + K q_{\perp}^4},\\
G^0_{zi}(\qv) &=& - \left({\frac{\rho B_z}{2\mu+\lambda}}\right)^{1\over2}
\frac{q_z q_i} {q_{\perp}^2 (B_z (1 - \rho)q_z^2 + K q_{\perp}^4)},
\hspace{1cm}\\
G^0_{ij}(\qv) &=& \frac{B_z q_z^2+K q_{\perp}^4}
{(B_z (1 - \rho)q_z^2
+ K q_{\perp}^4)} \frac{q_i q_j}{(2\mu+\lambda) q_{\perp}^4}\\
&+& \frac{1}{\mu q_{\perp}^4} (\delta_{ij} q_{\perp}^2 - q_i q_j),\nonumber
\end{eqnarray}
where $\rho = (B-\mu_L)^2/(B_z(2\mu+\lambda))$.  

To assess the role of nonlinear elastic terms in the presence of
thermal fluctuations we compute perturbative corrections to elastic
constants in ${\cal H}_{NE}$ and find that they are dominated by $u_z$
fluctuations.  Given the structure of $G^0_{zz}$ correlator, as in
smectic liquid crystals this perturbation theory diverges for $D\leq3$
on length scales longer than $\xi_{NL}^\perp=\Lambda^{-1}
g_\perp^{-1/(3-D)}$ ($=\Lambda^{-1}e^{32/(59g_\perp)}$, for $D=3$),
with $g_\perp$ defined in Eq.\ref{couplings_dimensionless}, below, and
$\Lambda$ an ultra-violet cutoff associated with the elastomer mesh
size.

To understand the elastomer properties beyond $\xi_{NL}$, we perform a
momentum-shell renormalization group calculation, perturbatively in
$\epsilon = 3-D$.  Integrating out short scale phonon fluctuations
$u_\alpha$ and rescaling the spatial coordinates and fields, so as to
keep $\Lambda$ fixed, we arrive at the following flow equations for
the elastic constants at scale $\Lambda^{-1}e^\ell$:
\begin{eqnarray}
\frac{d B}{d\ell} &=& (D+3-3 \omega - \eta_B) B ,\nonumber\\
\frac{d C}{d\ell} &=& (D+3-3 \omega- \eta_C) C ,\nonumber\\
\frac{d \mu_L}{d\ell} &=&  (D+3-3 \omega- \eta_L) \mu_L,\nonumber\\
\frac{d \mu}{d\ell} &=& (D+3-3 \omega - \eta_{\perp}) \mu,\nonumber\\
\frac{d K}{d\ell} &=& (D-1-\omega+\eta_K) K,
\label{couplingFlows}
\end{eqnarray}
where,
\begin{eqnarray}
\eta_B&=& \rho_2^2 g_L,\hspace{3mm} 
\eta_C=\eta_L = g_L,\hspace{3mm}
\eta_{\perp} = \frac{1}{8} g_{\perp},\\
\eta_K &=&\frac{{g_L}{g_{\perp}}
       \left( 9 + 17{{\rho }_1} -
         22{\sqrt{{{\rho }_1}}}{{\rho }_2} \right)  -
      4{g_L}^2 \left( -1 + {{{\rho }_2}}^2 \right) }
    {8\,\left( {g_{\perp}}\,{{\rho }_1} +
      {g_L}\,\left( 1 + {{\rho }_1} -
         2\,{\sqrt{{{\rho }_1}}}\,{{\rho }_2} \right)  \right) }.
\nonumber
\end{eqnarray}

The physics is controlled by the flows of two dimensionless coupling,
$g_L(\ell)$ and $g_{\perp}(\ell)$, and two dimensionless ratios
$\rho_1(\ell)$ and $\rho_2(\ell)$:
\begin{eqnarray}
g_\perp &=&\frac{\mu}{4\pi K^{\frac{3}{2}}} \,{\sqrt{\frac{
          \left( B - 2\,C + \mu  + {{\mu }_L} \right) }{B\,\mu  +
          2\,C\,\left( -2\,C + \mu  \right)  +
          \left( 4\,B + \mu  \right) \,{{\mu }_L}}}},\nonumber\\
g_{L} &=& {\mu_L\over\mu} g_\perp,\hspace{3mm}
\rho_1 = \frac{\mu_L}{B},\hspace{3mm}
\rho_2 = \frac{C}{\sqrt{B \mu_L}}.\label{couplings_dimensionless}
\end{eqnarray}
Flow equations for $\rho_1(\ell)$ and $\rho_2(\ell)$ are given by:
\begin{eqnarray}
\frac{d\rho_1}{d\ell} &=& - g_L \,\rho_1 \,(1 - \rho_2^2),
\label{rho1flow}\\
\frac{d\rho_2}{d\ell} &=& -\frac{1}{2} g_L\, \rho_2\, (1 -
\rho_2^2),\label{rho2flow}
\end{eqnarray}
where mechanical stability requires $|\rho_2| < 1$. With the exception
of an unstable fixed point at $g_L^*=0$ and an unstable fixed line at
$|\rho_2|=1$ (parameterized by $\rho_1$)\cite{XRunpublished}, we
expect (and verify a posteriori) that the infrared stable fixed point
is characterized by $g_L^*\neq 0$. Equations
\ref{rho1flow},\ref{rho2flow} then imply that at such fixed point,
$\rho_1^*=\rho_2^*=0$. The flow equations for $g_L(\ell)$ and
$g_\perp(\ell)$, the full form of which is too involved to be
reproduced here\cite{XRunpublished}, then simplify considerably:
\begin{eqnarray}
\frac{d g_L}{d\ell} &=& \epsilon g_L
 - \frac{{g_L}\left( 40{{g_L}}^2 + 68{g_L}{g_{\perp}} +
      13{{g_{\perp}}}^2 \right) }{8
    \left( 4{g_L} + {g_{\perp}} \right) },\label{flow_gL}\\
 \frac{d g_{\perp}}{d\ell} &=& \epsilon g_\perp
- \frac{ {g_{\perp}}\left( 4{{g_L}}^2 + 32{g_L}{g_{\perp}} +
        7{{g_{\perp}}}^2 \right) }{4
    \left( 4{g_L} + {g_{\perp}} \right) }.\qquad\label{flow_gperp}
\end{eqnarray}
In addition to reproducing the $3-\epsilon$-dimensional thermal
smectic fixed point\cite{GP}, with $g_\perp^*=0, g_L^*=4\epsilon/5$,
we find an infrared stable fixed point at $g_\perp^*=32\epsilon/59,
g_L^*=4\epsilon/59$, with
\begin{eqnarray}
\eta_B&=&0,\hspace{3mm} \eta_K=38\epsilon/59,\label{etas1}\\
\eta_C&=&\eta_L=\eta_L=\eta_\perp=4\epsilon/59,\label{etas2}
\end{eqnarray}
characterizing long-scale anomalous elasticity of homogeneous nematic
elastomers\cite{XRunpublished}.

For the physically most interesting case of $D=3$, $g_L(\ell)$ and
$g_\perp(\ell)$ are marginally irrelevant, allowing an asymptotically
{\em exact} computation of anomalous elasticity. Standard asymptotic
analysis of Eqs.\ref{rho1flow}-\ref{flow_gperp} with $\epsilon=0$,
shows that for large $\ell$ coupling constants flow to zero
algebraically:
\begin{eqnarray}
\rho_1(\ell)&\sim&\ell^{-4/59},\label{rho1}\\ 
\rho_2(\ell)&\sim&\ell^{-2/59},\label{rho2}\\ 
g_L(\ell)&\approx &\frac{1}{8}g_{\perp}(\ell)\approx\frac{4}{59\ell}.
\label{g_flows3D}
\end{eqnarray}

Matching calculations, together with Eqs.\ref{couplingFlows} then
predict that on long scales all effective elastic constants except $B$
are anomalous, depending logarithmically on wavevector $\qv$ of the
deformation
\begin{eqnarray}
B(\qv) &\cong& B_0,\\
\mu_L(\qv)\sim\mu(\qv)\sim C(\qv) &\propto&
(\log|\qv_{\perp}|)^{-4/59},\\
K(\qv)  &\propto& (\log|\qv_{\perp}|)^{38/59}.
\end{eqnarray}
Hence at long scales, the bulk modulus is much larger than shear
moduli, predicting that independent of microscopic details nematic
elastomer is effectively incompressible in the thermodynamic limit.
Logarithmically divergent strain-stress response immediately follows.
Equations \ref{couplings_dimensionless}, \ref{g_flows3D} predict a
universal ratio of the renormalized shear moduli
\begin{equation}
{\mu^R\over\mu^R_L}=8.\label{ratio}
\end{equation}
Although detecting logarithmic length scale dependence predicted here
is likely to be difficult, the incompressibility of nematic elastomer
and above universal ratio should be readily observable.

Based on our predictions in $3-\epsilon$ dimensions, we would expect a
much stronger, power-law anomalous elasticity in two-dimensional (2D)
nematic elastomers, with anomalous exponents given by
Eq.\ref{etas1},\ref{etas2} with $\epsilon=1$. However, a moment of
reflection shows that because in 2D the subspace
perpendicular to the nematic direction is one-dimensional, there is no
transverse shear mode. We must therefore set $\mu =0$, or 
deduce the properties 2D elastomers from a $D-1$-axial
analytical continuation, as we have recently done for nematic
elastomer membranes\cite{membranePRE}. A more serious difficulty is
that in our analysis we have ignored the subdominant columnar-like
elastic nonlinearities, that, although irrelevant near $D=3$, are
relevant for $D<5/2$ and in two dimensions are in fact identical to
the smectic-like nonlinearities that we have studied here.
Simultaneously treating both nonlinearities remains an interesting and
challenging problem.

Throughout this work, we have also ignored elastomer heterogeneities.
However, real elastomers are only {\em statistically} homogeneous and
isotropic, and therefore results presented here are only valid up to a
long but finite disorder-dependent length scale $\xi_{NL}^\Delta$. On
longer scales network heterogeneity dominates over thermal
fluctuations and results presented here crossover to a distinct
phenomenology, controlled by a nontrivial zero-temperature fixed
point\cite{XRdisorderPRL,RT_smectic,RT_columnar,RT_MSC}.

To summarize, we have presented a unified formulation of a
rotationally invariant nonlinear elasticity for a variety of elastic
media with spontaneously broken spatial symmetry. We applied this
formalism to nematic elastomers and studied their thermal
fluctuations. We found that they exhibit anomalous
elasticity, which in three dimensions is characterized by a
logarithmic wavevector dependence of elastic moduli, universal ratio
of shear moduli, and a long-scale incompressibility.

We thank T. Lubensky, O. Stenull and J. Toner for discussion and acknowledge
support by the NSF MRSEC DMR-0213918 (LR, XX), the 
Packard Foundation (LR), and the University of Colorado Faculty
Fellowship (LR), as well as KITP Graduate Fellowship through the NSF
PHY99-07949 (XX).  We thank Harvard Department of Physics, where
part of this work was done, for hospitality.


\begin{thebibliography}{99}
\bibitem{semisoft} Here we focus exclusively on ``ideal'' gels that
  exhibit a statistically isotropic and homogeneous high-temperature
  phase, in contrast to the so-called ``semi-soft'' gels that are
  intrinsically anisotropic.
\end{thebibliography}
\end{document}